# Three-Dimensional Atomic-Scale Structural Transformation in a $SrTiO_3$ Grain Boundary

**Authors:** Xiaoyue Gao[1#], Jiake Wei[2#], Bo Han[1], Junpin Luo[1], Ruilin Mao[1], Xiaowen Zhang[1], Xiaomei Li[3], Ryo Ishikawa[4], Bin Feng[4]*, Naoya Shibata[4], Yuichi Ikuhara[4]* and Peng Gao[1,2,5]*

**Affiliations:**

[1]International Center for Quantum Materials, and Electron Microscopy Laboratory, School of Physics, Peking University, Beijing 100871, China.

[2] Tsientang Institute for Advanced Study, Zhejiang 310024, China.

[3] School of Integrated Circuits, East China Normal University, Shanghai, 200241, China.

[4] Institute of Engineering Innovation, School of Engineering, The University of Tokyo, Tokyo 113-8656, Japan.

[5] Hefei National Laboratory, Anhui 230088, China.

[#]Equal contribution.

*Corresponding author. Email: feng@sigma.t.u-tokyo.ac.jp; ikuhara@g.ecc.u-tokyo.ac.jp; pgao@pku.edu.cn

**Abstract**

Grain boundaries (GBs) in complex oxides play critical roles in governing their functional properties, which are intrinsically linked to their three-dimensional (3D) atomic configurations and local chemical environments that can deviate markedly from those of the bulk. However, the 3D atomic structures of GBs remain poorly understood due to the projection limitations of conventional (S)TEM. Here, using multislice electron ptychography, we resolve the 3D atomic structure of a Σ13(510)/[001] tilt GB in $SrTiO_3$ with simultaneous visualization of both cation and oxygen columns. Depth-resolved reconstruction reveals pronounced structural inhomogeneity along the GB, uncovering a transition from the canonical symmetric configuration (STR1) to an asymmetric configuration (STR2) that is hidden in conventional projection imaging. Quantitative analysis of atomic-column intensities demonstrates that these two GB configurations possess distinct local chemical and vacancy distributions. By further mapping the atomic displacement fields, we reveal that the transformation between STR1 and STR2 proceeds via local atomic shuffling at the GB core and collective shear displacement in the adjoining grains, mediated by the step and dislocation character of the junction, respectively. Moreover, analysis of oxygen octahedral rotations reveals a strong dependence on the local atomic structure with pronounced asymmetry around the STR2 region. These findings establish a direct link among the 3D atomic structure, local chemical composition, and lattice order parameters at the GB, underscoring the critical importance of depth-resolved characterization in understanding and engineering GB-mediated functionalities in complex oxides.

## Introduction

Grain boundaries (GBs) are fundamental structural defects in polycrystalline materials, where the atomic configurations differ markedly from the surrounding bulk. As intrinsically three-dimensional (3D) interfacial systems, GBs can host complex atomic reconstructions, non-uniform chemical distributions, and coupled lattice distortions, giving rise to a rich landscape of metastable structural states. Increasing experimental and theoretical evidence has established that GBs constitute bulk-supported interfacial phases, capable of adopting multiple distinct structural states and undergoing phase transformations in response to changes in temperature, chemical environment, and external fields[1-7]. Such GB structural complexity and phase behavior play a decisive role in governing material properties that are dominated by interfaces[8-11]. For example, in complex oxides such as $SrTiO_3$ (STO), GBs have been shown to induce emergent phenomena including flexoelectricity[12] and room-temperature antiferrodistortive (AFD) phase transitions[13], while also strongly modulating dielectric response and charge transport through interfacial charge redistribution[14-18]. In face-centered cubic metals, variations in local composition and temperature at GBs can trigger abrupt GB phase transitions, leading to discontinuous changes in GB diffusivity and improved irradiation resistance[2-5]. These findings suggest that understanding the 3D atomic structures of GBs is central to revealing the origin of their interfacial phase behavior and functionalities.

Despite their importance, direct experimental determination of the atomic GB structures at the 3D remains highly challenging. Although advances in scanning transmission electron microscopy (STEM) have enabled atomic-scale characterization of GB structures and properties such as the structural unit, strain[19], polarization[20, 21], elemental[17, 22] and charge distributions[14, 16, 17], conventional STEM is intrinsically restricted to two-dimensional (2D) projection imaging. As a result, the complex 3D atomic configurations, defect distributions, and order-parameter variations at GBs collapsed into averaged contrasts, obscuring depth-dependent structural heterogeneity. This limitation is particularly severe for oxide GBs, where the lattice distortions, oxygen octahedral rotations (OORs), and vacancy segregation along the projection direction critically influences the GB structural stability and functionality. Moreover, electron channeling in conventional STEM imaging further complicates the quantitative image interpretation, making it difficult to unambiguously determine the atomic occupancy and stoichiometry near the GBs[22-24]. Recent advances in multislice electron ptychography (MEP) provide a promising route to overcome these limitations[25-28]. By combining four-dimensional STEM datasets with advanced phase-retrieval algorithms, MEP enables quantitative 3D atomic structure reconstruction with deep sub-ångström resolution in the lateral directions, nanometer-scale resolution along the depth direction and enhanced sensitivity to vacancies[26, 29-33], effectively circumventing the projection constraints of conventional STEM.

In this work, we employ the MEP to quantitatively resolve the 3D atomic structure of a Σ13(510)/[001] tilt GB in STO and characterize the atomic scale inhomogeneities that govern its functional properties. We reveal that the GB undergoes a depth-dependent phase transition, evolving from the symmetric Σ13 configuration (STR1) to an asymmetric configuration (STR2), although conventional projection imaging methods only identify the canonical GB structure. By quantitatively resolving the atomic-column intensities of all constituent species (O, Sr, and TiO) across the GB, we demonstrate that the two GB configurations possess different local chemical compositions. Moreover, the GB structural transformation proceeds via local atomic shuffling coupled with the collective shear displacement mediated by the step and dislocation character respectively of the junction between STR1 and STR2. Quantitative analysis further reveals that OORs near the GB are strongly correlated with the local GB configurations, exhibiting pronounced anisotropy in both in-plane and out-of-plane rotational components, which implies that the two GB structures should exhibit different properties (e.g. the electrical conductivity[21], optical properties[34], and dielectric response[35]). These results establish a direct 3D atomic-scale link between interfacial phase behavior, compositional inhomogeneity, and local properties, providing a general framework for understanding and engineering GB functionality in complex oxides.

**Results**

Figure 1a shows a high-angle annular dark-field (HAADF) STEM image of the Σ13(510)/[001] tilt GB in STO, which resolves the projected positions of the cation sublattice. In contrast, the projected MEP reconstruction (Fig. 1b) provides enhanced spatial resolution and, crucially, enables the direct visualization of oxygen columns, allowing a complete atomic-scale description of the GB structure. By integrating these results with the atomic resolution chemical sensitive energy-dispersive X-ray spectroscopy maps (Supplementary Fig. S1), we constructed the GB structural unit shown in Fig. 1c, where all the Sr, TiO and O atomic columns are included. The resulting projected GB structure forms a symmetric structural unit, which is consistent with previous results[13, 36]. When analyzing the depth-resolved structural information from the MEP reconstructed image, we find that this GB exhibits pronounced inhomogeneity along the depth direction. As illustrated in Fig. 1d, several GB atomic columns exhibit depth-dependent lateral offsets. Quantitively analysis of the atomic positions at different depths (Fig. 1e) further indicates that this structural deviation is primarily localized within a depth range of approximately 13-16 nm. These results reveal that the GB could have depth-dependent structural distortion and 3D structural complexity, which could be obscured by projection-based analyses.

To elucidate the origin of the observed depth-dependent structural deviation, we divided the GB into three regions along the depth direction and analyzed their GB structures individually in Figs. 2a-c. In the upper region, the GB (Fig. 2a and Fig. S2) maintains a symmetric structure, which is labeled as the structural model STR1 (Fig. 2d). This GB structure is the same as those observed in conventional HAADF-STEM image and the projected MEP shown in Figs. 1a-b. In the transition region (13-16 nm), the GB configuration (Figs. 2b and 2e) is likely caused by the overlap of the upper and lower GB structures due to the limited depth resolution of MEP. In contrast, the GBs change to a distinctly asymmetric structure (dotted red lines in Figs. 2c and 2f) in the lower region, which is remarkably different from the STR1, indicating a depth-dependent structural transition along the GB. Compared to STR1, STR2 is slightly shifted to the left side, as indicated by the red arrow in Fig. 2c, leading to a transformation from Sr-O-terminated STR1 (Fig. 2d) to TiO-O termination (Fig. 2f). This GB movement is equivalent to the rigid body shift of the two adjoint grains and changes the local GB symmetry, resulting in a symmetrical to unsymmetrical GB structural transformation. To further probe the atomic-scale response within the GB core, we compared the behavior of atomic sites along the depth direction. As shown in Fig. 2g, the oxygen columns exhibit pronounced variations in both position and intensity within the 13-16 nm transition region. The apparent elongation of atomic columns in the projected image during the transition region can be attributed to noticeable variations in position and intensity of some TiO sites (e.g., Columns 16 and 17 in Fig. S3) in the core. In contrast, the Sr columns (Columns 2-7, Fig. 2h) exhibit relatively stable positions in the transition region. This pronounced disparity shows that different atomic sublattices within the GB exhibit distinct responses to the depth-dependent GB phase transition.

To elucidate the chemical composition variation associated with the observed GB structural transformation, we quantitatively analyze atomic-column intensities at the GB using ptychographic reconstruction. Unlike conventional imaging techniques that are strongly affected by electron channeling effect and thus limited in quantitative atomic occupation at distorted defect cores[23], MEP enables a near-linear relationship between the image intensity and atomic occupation[25, 26], allowing direct assessment of stoichiometry and vacancy distributions within the GB. As shown in Figs. 3a-c, the normalized phase maps of the three types of atomic columns (Sr, O, and TiO) within the STR1 GB core all exhibit a pronounced reduction relative to the bulk reference, indicating a general depletion of atomic occupancy and the presence of vacancies in the GB region. These observations are also supported by the electron energy-loss spectroscopy results shown in Fig. S6, which suggests the existence of oxygen-vacancy accumulation in the GB. Notably, this intensity suppression is not spatially uniform. While Sr columns located near the coincidence-site lattice positions of the two grains (Columns 1 and 7, as indicated by arrows in Fig. 3a) retain intensities comparable to those in the bulk, most other Sr sites within the GB core show a substantial decrease (Fig. 3a). This spatially heterogeneous contrast suggests that the GB does not simply undergo a uniform atomic depletion, but instead exhibits site-dependent vacancy formation that is likely governed by the local structural constraints and coordination environments within the boundary core. Furthermore, a clear asymmetry in Sr intensity is observed between the left (yellow regions) and right (blue regions) side of the STR1 GB (Fig. 3a and Fig. S4), with the left side consistently exhibiting lower intensity than the right side. These results indicate an uneven re-distribution of vacancies across the GB. Similar asymmetric distribution is also observed for O (Fig. 3b) and TiO sites (Fig. 3c) in STR1 although this GB shows a symmetric structural unit.

The reduced intensities of the atomic columns in the core of STR2 also indicate the persistence of vacancies within this asymmetric GB configuration (Fig. 3d-f). Meanwhile, there are some columns exhibits pronounced intensity variations with the corresponding columns in STR1. For example, the Sr intensity of Columns 8-12 (blue regions in Figs. 3a and 3d) in STR1 increases with the GB structural transition to STR2, whereas the intensity of Columns in yellow regions decreases. As shown in Figs. 3c and 3f, the TiO intensities in Columns 17 and 23 (marked by red and black dashed circles) also exhibit pronounced changes during the phase transition, with the intensity increasing in Column 17 while decreasing in Column 23. To further quantify these chemical variations, we analyzed the intensity line profiles of the Sr, O, and TiO sites (Figs. 3g-i), providing a more direct comparison between STR1 and STR2 across the GB. Notably, the intensity of Sr columns on the left side of the GB shows a pronounced reduction in STR2 compared to STR1 (red shadow in Fig. 3g), indicating a higher concentration of Sr vacancies in STR2. The intensity of O columns in the STR2 exhibits clear increases (orange shadow in Fig. 3h). Meanwhile, the intensity of TiO columns on the left side also decreases in STR2 compared to STR1 (green shadow in Fig. 3i). Collectively, these results suggest that the GB phase transition is closely coupled to vacancy redistributions and local defect accumulations or absorptions. It is important to note that, the intensities of specific O sites in the STR2 structure are markedly enhanced ( marked by black arrows in Fig. 3e) and abnormally exceed half of the bulk intensity, which are also highlighted by black circles in Fig. 2f. We attribute this phenomenon to mixed occupancy, where metal and oxygen ions may potentially coexist or intermix with different occupation depths at these specific GB coordinates in the STR2 phase.

To gain deeper insight into how the GB structure evolves from STR1 to STR2, we superimpose the atomic positions of the two structures, as shown in Fig. 4a. The comparison shows that the atomic positions in STR1 do not fully coincide with those in STR2. Such mismatch is not only observed at the GB core but also extends into the adjacent bulk-like regions. To quantify the atomic displacements, we map the difference vectors of the Sr columns from their positions in STR1 to those in STR2, as illustrated by the red arrows in Fig. 4b. These vectors might be interpreted as the atomic displacements associated with the transformation of the GB structure from STR1 to STR2 if ignoring the atom occupation variations of each column. In the GB region, the relatively large displacement suggest pronounced local atomic shuffling with the structural transformation. This shuffling-like motion also effectively leads to the lateral migration of the GB toward the left side during the STR1-to-STR2 transition[37], which results in a step character at the junction between STR1 and STR2. In contrast, in the bulk region, the atomic displacements are more uniform and collective. The collective nature of the bulk displacement field implies a rigid-body-like shift of the adjoining grains during the STR1-to-STR2 transition, suggesting that the STR1/STR2 junction exhibits a dislocation-like character[2, 3, 6, 38]. The Burgers vector of the junction is calculated to be the displacement difference vector by the construction of the Burgers circuit as the pink lines in Fig. 4c. The displacement vectors of the left and right grains are averaged and represent as vectors AA' (DD') and BB'(CC') respectively, where A, B, C, D and A', B', C', D' represent the atomic columns positions in the STR1 and STR2, respectively. The size of the Burgers vector is calculated to be ~21 pm, which is much smaller than the conventional lattice dislocations in STO.

Building on this structural framework, distinct GB structural configurations are expected to exhibit different local order parameters and, consequently, distinct functional properties. To further clarify the structure-property implications of the depth-dependent GB phase transition, we quantitatively analyzed the OORs across the GB based on Fig. 4a. As a fundamental structural degree of freedom in perovskite oxides, OORs dictate the AFD phase transitions[13, 34, 39], local symmetry breaking[40, 41], and a wealth of emergent functionalities including dielectric anomalies[13, 42], photoelectronic responses[43, 44] and transport anisotropy[45-47]. The in-plane OORs component $\theta_v$ (Fig. 5a) exhibits pronounced rotation primarily localized at the edges of the GB core, with magnitudes significantly larger than those of out-of-plane components $\theta_h$. In contrast, the $\theta_h$ component (Fig. 5b) remains relatively weak and more uniformly distributed across the GB region. These characteristics of the OORs distribution in the symmetric STR1 configuration are consistent with our previous experimental observations and prior theoretical calculations[13, 48]. In striking contrast, the OORs states of the STR2 (Figs. 5c and 5d) differ distinctly from those of STR1. The rotation angles in both directions exhibit significant asymmetry, reflected by ~2° larger OORs on the left side. In addition, STR2 exhibits prominent $\theta_h$ around the GB as shown in Fig. 5d, whose magnitudes are even larger than the $\theta_v$ of STR2. To visualize these differences quantitatively, we compared the rotation angles between STR1 and STR2 in Figs. 5e and 5f. The difference maps (Figs. 5e and 5f) provide a direct visualization of the redistribution of OORs during the GB phase transition. Both the out-of-plane component $\theta_v$ and the in-plane component $\theta_h$ exhibit pronounced spatial reorganization as the GB structure evolves along the depth. Specifically, lattices with large octahedral rotation angles in STR1 are progressively incorporated into the core region of the STR2 phase (Fig. 5e). Meanwhile, new regions with enhanced octahedral rotations emerge at the left side of STR2 (Fig. 5f). It indicates that the GB phase transition is accompanied by the magnitude and spatial distributions reconfiguration of OORs, rather than a simple preservation of the initial rotation states. These results demonstrate that the spatial distribution, orientation, and magnitude of OORs are strongly modulated by the GB phases.

**Discussion**

In this work, we uncovered a depth-dependent structural inhomogeneity in the Σ13 (510)/[001] tilt GB of STO, which is inaccessible via conventional projection imaging. By leveraging MEP, we identify two distinct GB configurations along the depth direction. Although previous experiments have already revealed the GB structural variations along the GB with projected STEM images[36, 49], our depth resolved results show this kind of inhomogeneity and GB structural variations could widely occur in GBs. Previous simulations have shown that a GB could adopt a wide variety of structures with comparable GB energy[3, 4, 23, 49]. It is reasonable that these stable/metastable GB structures could coexist within the bicrystal. However, these GB structural inhomogeneity and diversity may be overlooked by the conventional projected imaging approaches.

Meanwhile, the existence and distribution of local chemical compositions and vacancies are quantitatively analyzed in this work by MEP. The intensity reduction observed at the atomic columns within the GB core directly indicates the presence of vacancies, suggesting that the GB region deviates significantly from the stoichiometric bulk structure. Such variations in vacancy concentration in STO may further influence the local electronic structures, local order parameter and functional properties of the GB, as oxygen vacancies are closely related to phenomena such as two-dimensional electron gases[50], back-to-back Schottky barriers[51, 52] and optical transitions[34, 44].

Furthermore, the two GB configurations, STR1 and STR2, exhibit distinctly different local chemical compositions and vacancy distributions. These results are consistent with previous atomistic simulations, which suggest different GB structural configurations may have different equilibrium stoichiometries and defect concentrations[23, 34, 49, 53]. The experimentally observed chemical differences between STR1 and STR2 therefore suggest that the GB structural transformation and OORs are closely coupled to point-defect redistribution and local chemical reconstruction.

The GB structural transformation observed here may be mediated by the nucleation and movement of the GB phase junction with both step and dislocation characters. Although previous theoretical studies have predicted that such GB phase junctions may carry dislocation characters[38], experimental verification has remained challenging due to the extremely small Burgers vectors involved. Here, ptychographic reconstruction provides both high signal-to-noise ratio and effective correction of spatial drift, enabling reliable quantification of subtle atomic displacements and thus direct identification of the dislocation-like character of the junction. Meanwhile, the motion of such junction during GB structural transitions can be accompanied by macroscopic deformation. This suggests that GB structural transformation may serve as an important mechanism for plastic deformation during GB migration and GB sliding, particularly in nanocrystalline materials.

Moreover, these structural phase transitions cause changes of the GB OORs, a key order parameter in perovskite oxides. While the OORs around the symmetric STR1 configuration exhibit relatively balanced and localized patterns, the GB with STR2 phase modifies the OORs with pronounced asymmetry in both magnitude and spatial distribution. The change of OORs indicates a substantial reconstruction of the local bonding environment during the phase transition. These results indicate that GB phase transitions not only involve atomic rearrangement but also fundamentally alter lattice order parameters, which are closely linked to functional properties such as dielectric response and transport behaviors[18, 21, 35, 44, 47, 54].

In summary, this work reveals that GBs in complex oxides could exhibit depth-dependent structural and chemical inhomogeneity, which is inaccessible by conventional projection imaging. By combining MEP with quantitative atomic-scale analysis, we identify coexisting GB phases along the depth direction, resolve their compositional differences, and uncover the GB phase junction-mediated transformation mechanism between them. The strong coupling among GB phase behavior, vacancy distribution, and OORs further demonstrates that such structural transitions are intimately linked to local lattice distortions and functional properties. These findings provide a comprehensive three-dimensional picture of GB structure and its transformation, highlighting the critical role of depth-resolved characterization in understanding and engineering GB-mediated functionalities in complex oxides.

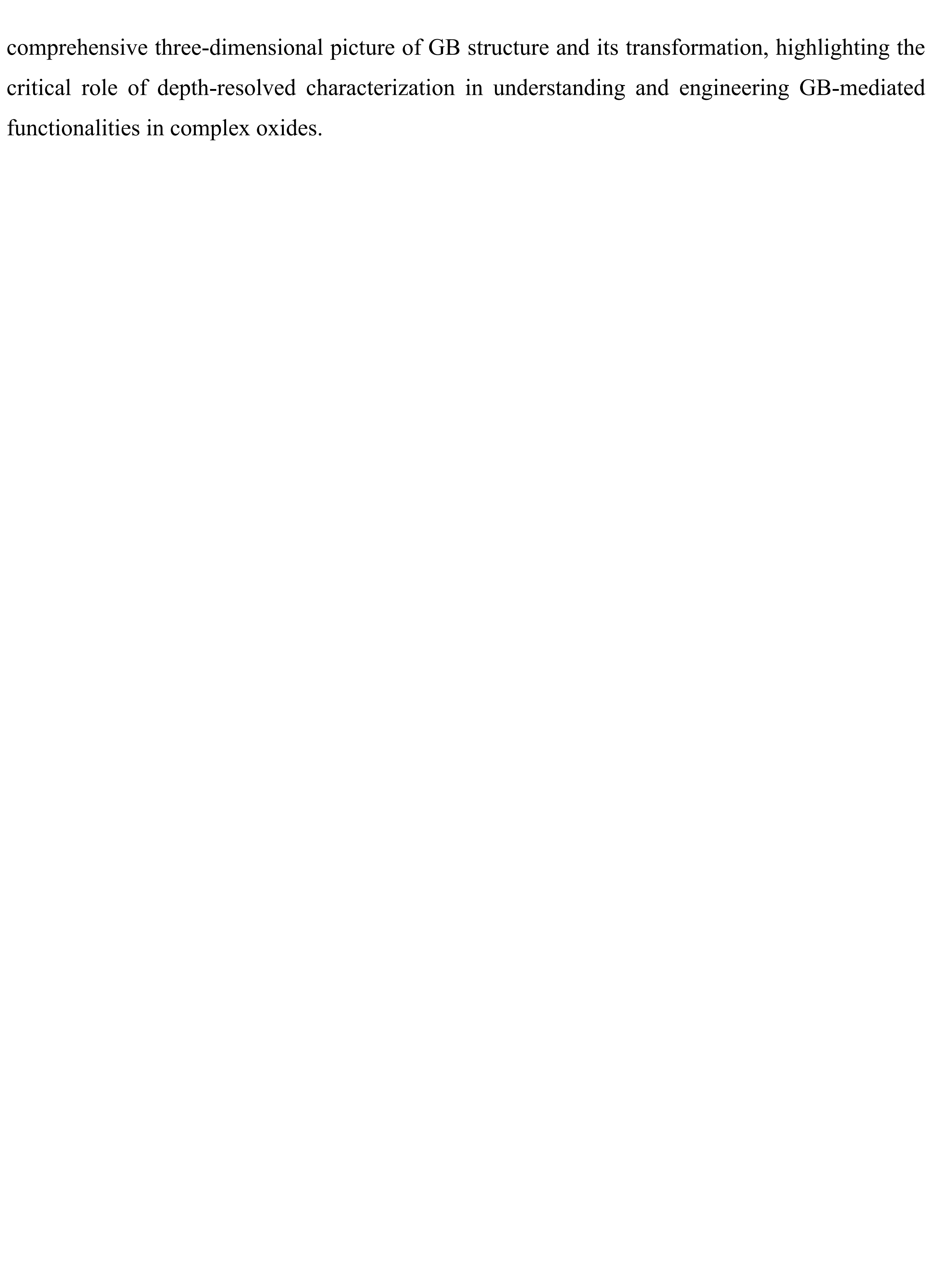

**Acknowledgments**

This work was supported by the National Natural Science Foundation of China (52125307 to P.G.), Grant-in-Aid for ScientificResearch (S) (JP22H04960), Grant-in-Aid for Scientific Research (B) (JP25K01522) and Grant-in-Aid for Scientific Research (A) (JP25H00793) from the Japan Society for the Promotion of Science (JSPS). P.G. acknowledges the support from the New Cornerstone Science Foundation through the XPLORER PRIZE. B.F. acknowledges the support from JST-PRESTO (Grant JPMJPR23JB), Japan. N.S. acknowledges the support from JST-ERATO (JPMJER2202), Japan. We acknowledge Electron Microscopy Laboratory of Peking University for the use of electron microscopes, and the High-performance Computing Platform of Peking University for providing computational resources. We thank Dr. jiade Li for helpful discussion.

**Author contributions:** X.Y.G. performed the STEM experiments and reconstructions. P.G. conceived the project; X.Y.G. performed the ptychographic experiments, reconstruction and data analyses with the assistance of B.H., J.P.L., R.L.M., X.M.L., R. I., B. F. and N.S.; B.H. and X.W.Z. performed the STEM-EELS experiment and analysis. X.Y.G. and J.K.W. wrote the manuscript under the direction of B. F., Y.I. and P.G.; All the authors contributed to this work through useful discussion and/or comments to the manuscript.

**Competing interests:** The authors declare no competing interests.

**Data and materials availability:** All data needed to evaluate the conclusions in the paper are available in the main text or the supplementary materials. Additional data and code related to this paper may be requested from the authors.

## SUPPLEMENTARY MATERIALS

Materials and Methods

Figs. S1 to S6

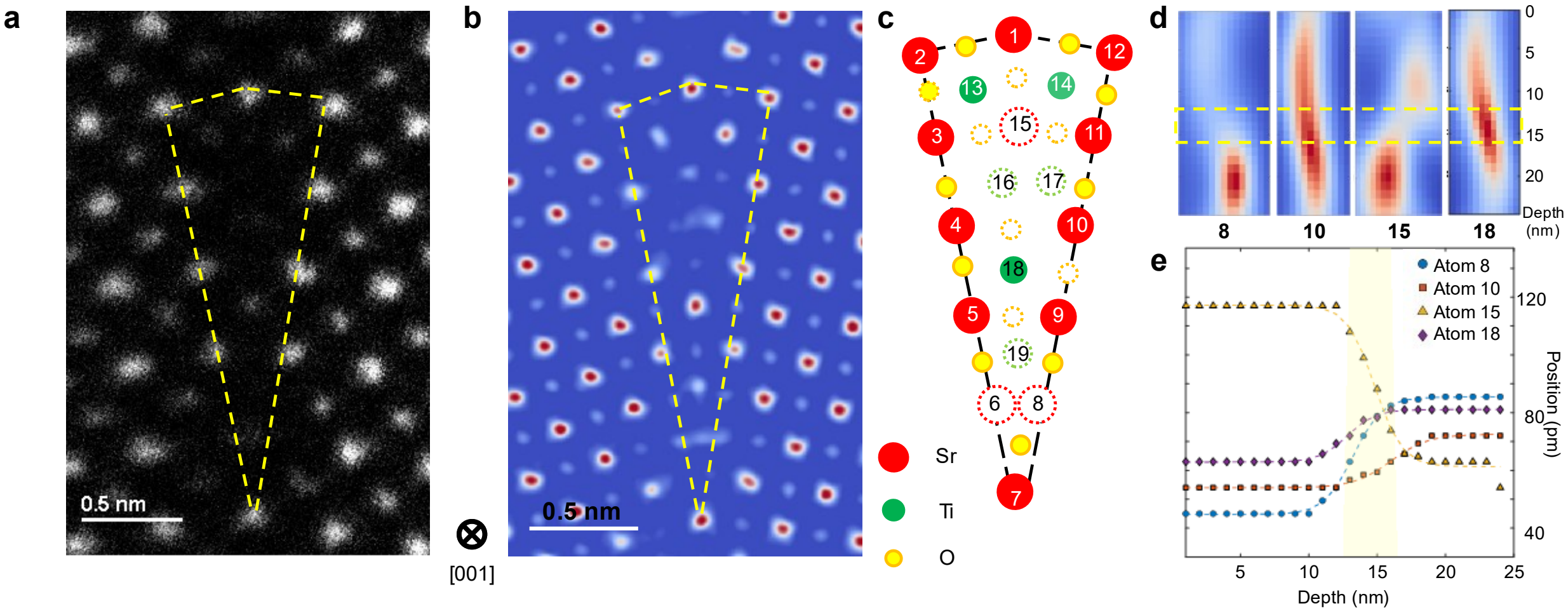

**Fig. 1. MEP imaging of the Σ13(510)/[001] STO GB. a,** The HAADF and **b,** MEP reconstructed projected image of the GB. **c,** The corresponding atomic structural unit model of the GB in a. **d,** Depth sectioning of some representative atoms. The depth region marked with yellow dashed line is the transition region marked in e. **e,** Linecuts of the corresponding atomic position in d, where the yellow shadow represents the depth region of the atomic transition.

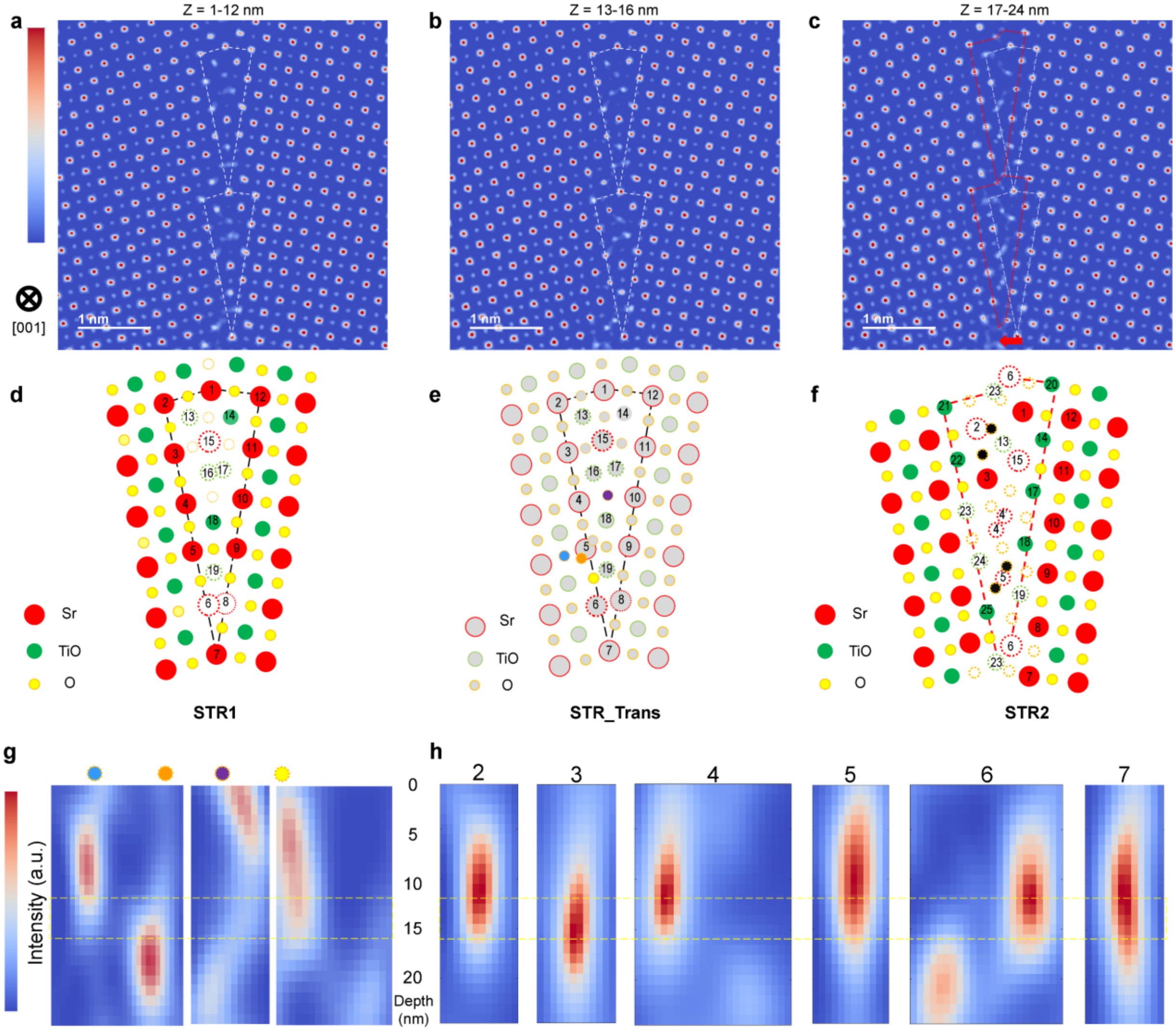

**Fig. 2. Structure inhomogeneity along the viewing direction. a-c,** Projected images with different depth ranges, i.e., 1-12 nm, 13-16 nm, and 17-24 nm respectively. The dotted white and red lines are the structural units of the upper and lower GB, respectively. The red arrow in (c) indicates the slight GB migration. **d-f,** The corresponding atomic GB models (named STR1, STR_Trans, STR2). The dashed atom columns mean significantly reduced intensities. **g,** The depth sectioning of the oxygen atom columns as marked in (e). **h,** The depth sectioning of atom columns No.2-7 (right side of the GB). The yellow dashed lines outline the transition region. The red arrow in c marks the GB migration direction from STR1 to STR2.

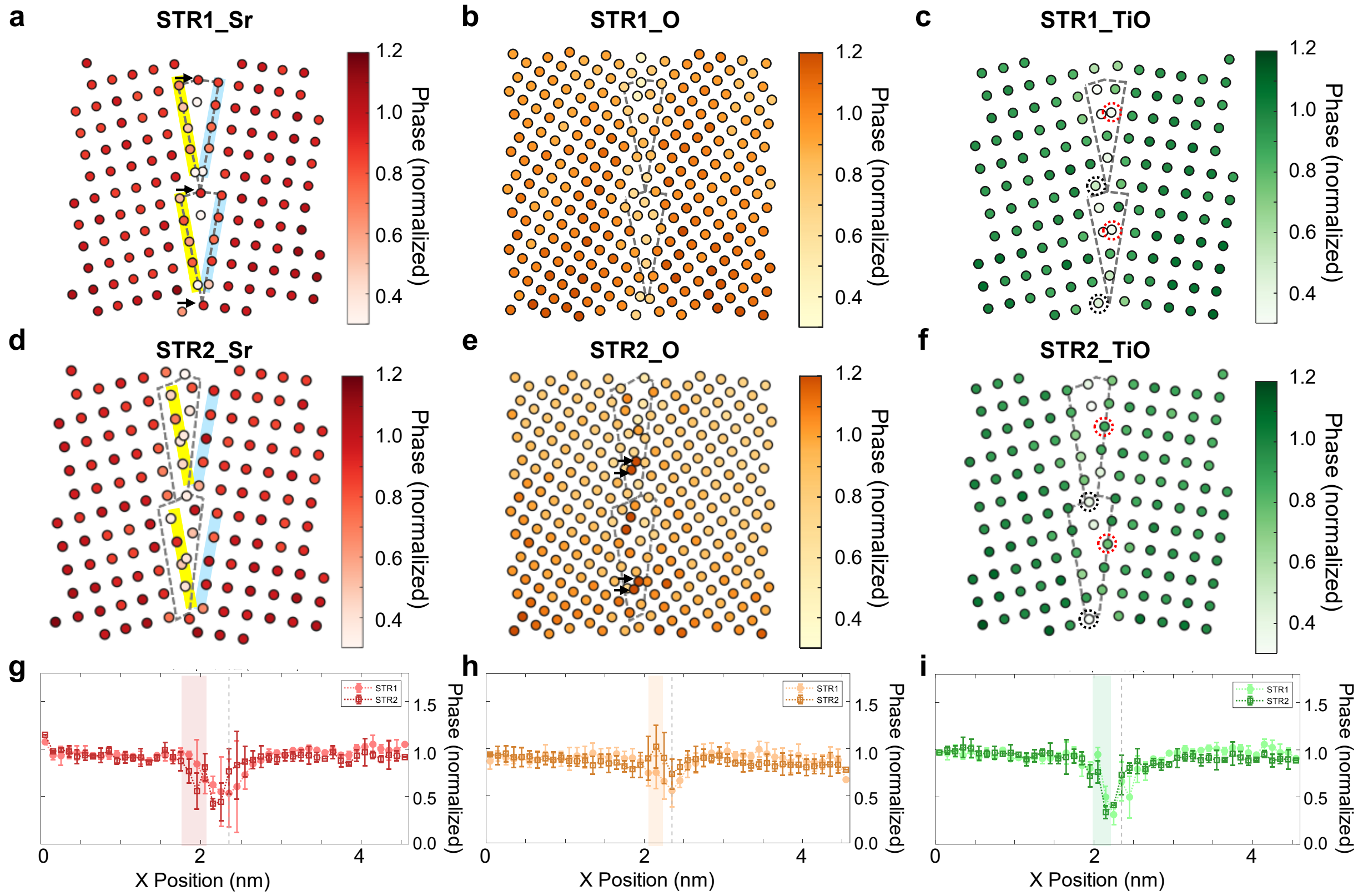

**Fig. 3. Atomic chemical composition analysis of STR1 and STR2. a-c,** Individually mean-bulk-normalized phase intensity maps of the Sr, O and TiO columns extracted from Fig. 2a with STR1, respectively. The black arrows in (a) mark Columns 1 and 7, which show similar intensity with the bulk columns. **d-f,** Normalized intensity maps of the Sr, O and TiO columns of Fig. 2c with STR2, respectively. The regions highlighted in yellow and blue indicate asymmetric intensity on the two sides of the STR1 (yellow, Columns 2-6; blue, Columns 8-12). The black arrows in (e) mark the abnormal high intensity sites. The red/black dashed circles in (c) and (f) mark the Columns 17/23, which show clear intensity change between STR1 and STR2. **g-i,** Intensity line profiles of Sr, O and TiO columns across the GB STR1 and STR2 (along the horizontal direction). Each data point averaged over 100 pm, with error bars denoting the standard deviation within each bin. The shaded regions highlight the intensity differences between STR1 and STR2 on the left side of the GB. The vertical dashed lines indicate the position of the GB core in STR1.

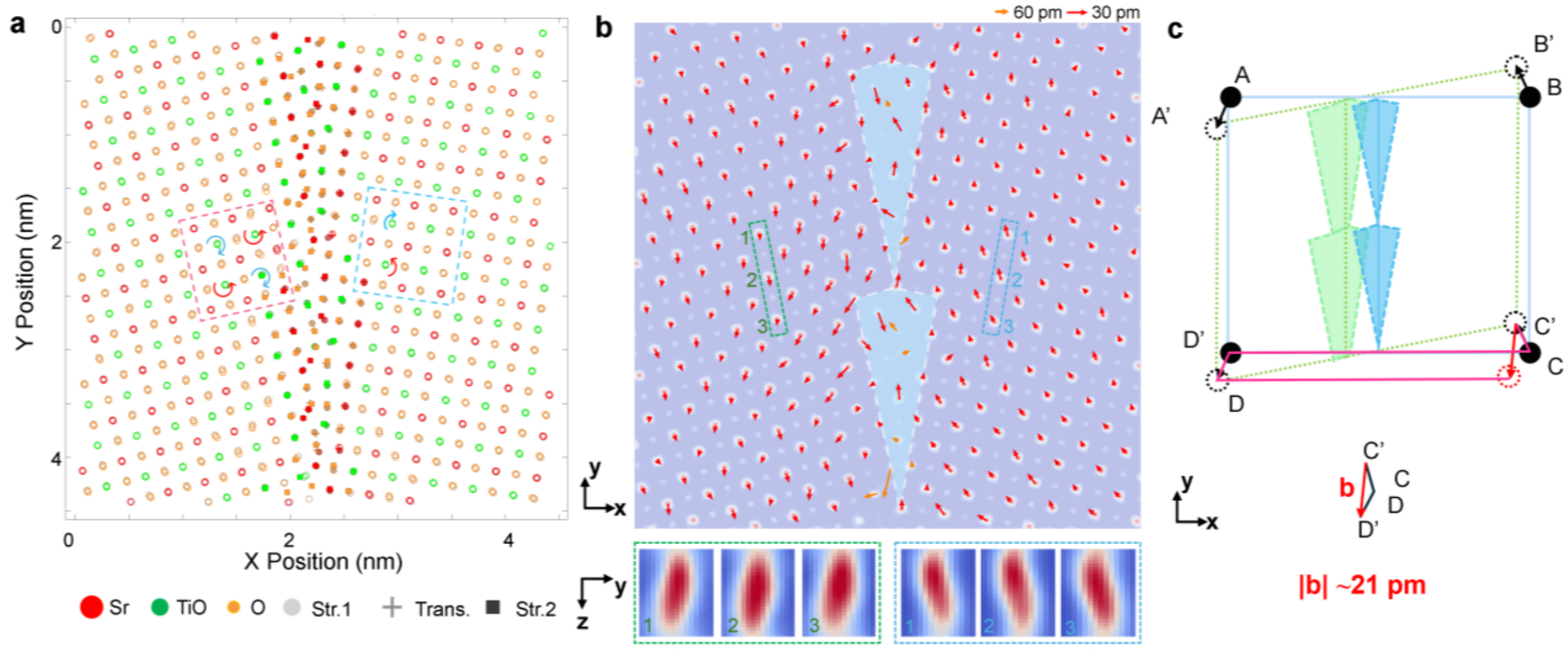

**Fig. 4. The distortion analysis of the GB along the depth direction. a,** Atomic position of different depth images (Fig. 2a-c), which are overlapped. Colors label different atom columns. Different symbols label three layers (STR1, STR_Trans and STR2). The red and blue arrows in the dashed boxes represent the counterclockwise and clockwise OOR along the depth direction, respectively. The dashed line boxes highlight the different degrees of OOR changes along the depth between the left and right grains. **b,** Displacement map of atomic positions overlaid on the MEP projected image. The arrows indicate the magnitude and direction of atomic displacements from the STR1 to STR2 in case of Sr. To clearly distinguish lattice distortions in the bulk and at the GB core, two sets of quiver scaling factors are applied: displacements smaller than 30 pm are shown as red arrows, displacements exceeding 30 pm are marked with orange arrows. STR1 structures are highlighted with light blue. The insets show the difference of the atomic position change between the left (green) and right (blue) sides of the GB, the arrows at the bottom indicate the opposite change along y direction. **c,** Schematic illustration of the 3D GB structure. The STR1 and STR2 with their adjoint bulk are overlapped and simplified shown as solid bule and dotted green lines respectively. The columns A, B, C, D and A', B', C', D' are corresponding to the bulk atomic columns in STR1 and STR2. The black arrows AA' and BB' correspond to the atomic displacement of the bulk atoms from STR1 to STR2. The Burgers vector is calculated as the difference of the displacement vector of the two grains, which is shown as the red arrow.

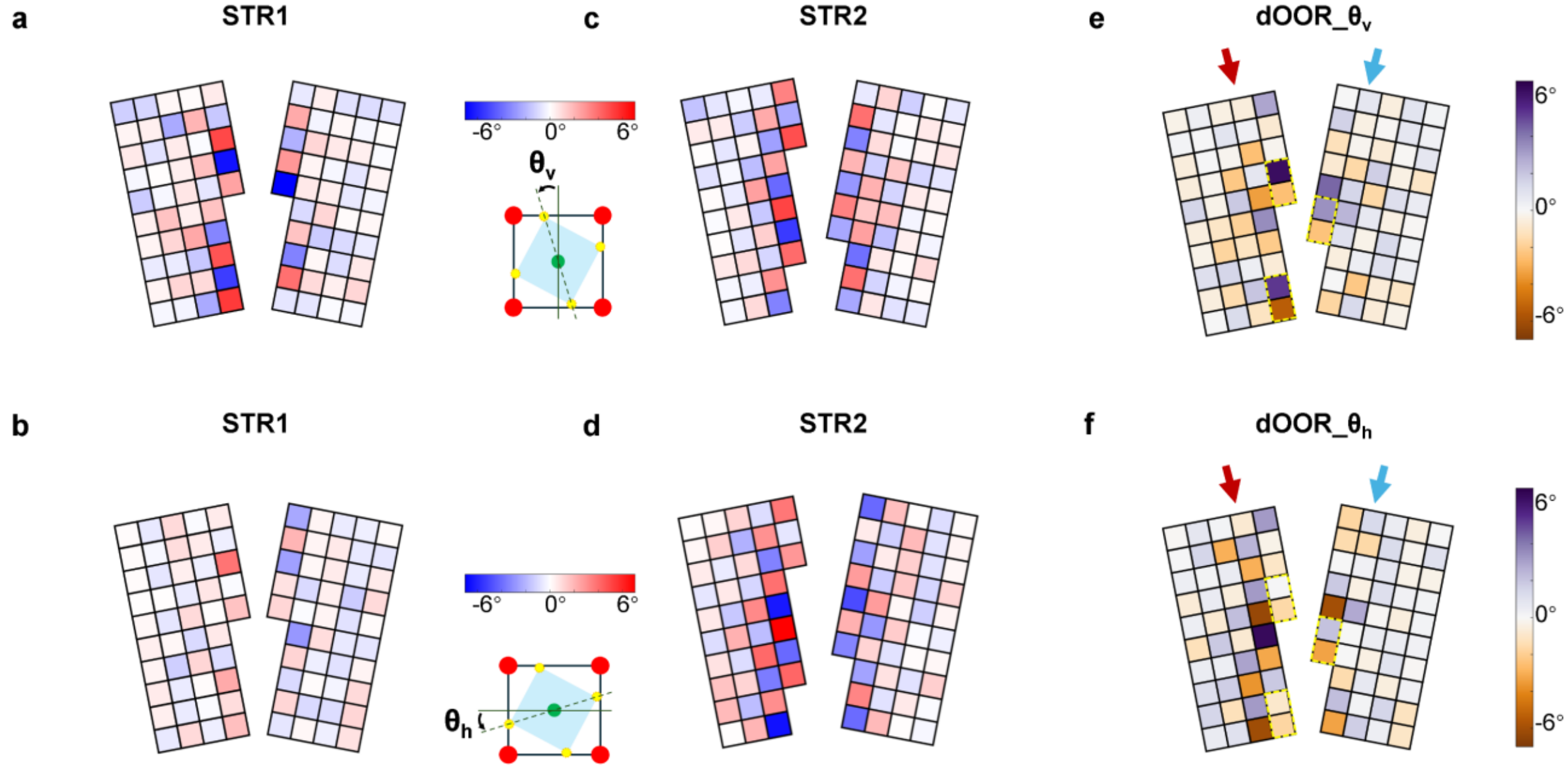

**Fig. 5. The oxygen octahedra rotations along the depth direction. a-b,** The extracted $\theta_v$ and $\theta_h$ maps around STR1 GB (Fig. 2a). **c-d,** The extracted $\theta_v$ and $\theta_h$ maps around STR2 GB (Fig. 2c). Red and blue represent counterclockwise and clockwise rotation of OOR, respectively. **e-f,** The differences of $\theta_v$ and $\theta_h$ between STR1 and STR2. Purple and orange indicate counterclockwise and clockwise rotation of STR2 relative to STR1, respectively. The yellow dashed boxes indicate that the OOR only exists in STR1 or STR2 due to the structure change.